\documentclass[a4paper,10pt,twoside]{cpc-hepnp}
\usepackage{CJK,upgreek,fancyhdr}
\usepackage{multicol}
\usepackage{graphicx}
\usepackage{booktabs}
\usepackage{amssymb,bm,mathrsfs,bbm,amscd}
\usepackage[tbtags]{amsmath}
\usepackage{lastpage}

\begin{document}
\begin{CJK*}{GB}{gbsn}

\fancyhead[co]{\normalsize
}
\footnotetext[0]{ }

\title{\large Hadronic coupling constants of $g_{\sigma\pi\pi}$ in lattice QCD
\thanks{
Supported by 
the National Magnetic Confinement Fusion Program of China (2013GB109000).
}
}
\author{
\quad Lingyun Wang$^{1}$
\quad Ziwen Fu$^{2,3}$\email{fuziwen@scu.edu.cn, corresponding author}
\quad Hang Chen$^{3}$
}
\maketitle

\address{%
$^1$International Affair Department, Chengdu Jiaxiang Foreign Language School, Chengdu 610023, China\date
}
\address{%
$^2$
Institute of Nuclear Science and Technology, Sichuan University, Chengdu 610064, P. R. China.
}
\address{%
$^3$Center for Theoretical Physics, College of Physical Science and Technology, Sichuan University,
Chengdu 610064, China
}

\begin{abstract}
We investigate the coupling constant $g_{\sigma\pi\pi}$ for the hadronic decay $\sigma\to\pi\pi$
only using the relevant three-point function,
which is evaluated by the moving-wall source technique with a pretty good noise-to-signal ratio.
This simulation is carried out on a $40^3\times96$ MILC gauge configuration with $N_f=2+1$ flavor
of the ``Asqtad'' improved staggered dynamical sea quarks at the lattice spacing $a \approx 0.09$ fm.
Our estimated value for this given MILC fine lattice gauge ensemble  $g_{\sigma\pi\pi}=2.71(42)$~GeV.
\end{abstract}


\begin{pacs}
{12.38.Gc,11.15.Ha}
\end{pacs}

\begin{multicols}{2}
\maketitle

\section{Introduction}
In 2016, Particle Data Group (PDG) lists $f_0(500)$,
which is often called $\sigma$ meson $I(J^{PC})=0(0^{++})$,
with mass $400-550$~MeV and broad width~$400-700$ MeV~\cite{Olive:2016xmw}.
Although direct determination of $\sigma$ resonance parameters
from QCD is difficult since it is a nonperturbative problem,
some theoretical efforts are still taken to study $\sigma$ meson and its resonance parameters~\cite{Pelaez:2015qba,Masjuan:2014psa,Oller:2003vf,Kaminski:2009qg,Mennessier:2010xg,GarciaMartin:2011jx,Pelaez:2010fj,Narison:2005wc,Nebreda:2010wv,Zhou:2004ms,Oller:1997ti,
Hyodo:2010jp,Caprini:2008fc,Yndurain:2007qm,Caprini:2005zr,
Escribano:2002iv,Giacosa:2007bn}.

The most feasible way to nonperturbatively obtain $\sigma$ resonance parameters
from first principles is using lattice QCD.
So far, there are just a couple of lattice reports on $\sigma$ resonance parameters.
The first lattice attempt 
is preliminarily performed on a MILC ``medium'' coarse lattice ensemble~\cite{Fu:2012gf}.
Nevertheless, the evaluation of vacuum diagram of $I=0$ $\pi\pi$ scattering
is not convincing~\cite{Fu:2012gf}.
The Hadron Spectrum Collaboration observed that the $I=0$ $\pi\pi$ scattering amplitude renders
the characteristic of a $\sigma$ looking as a broad resonance for the lighter quark mass,
which  very  resembles the experimental case~\cite{Briceno:2016mjc}.
The ETM Collaboration study with the $s$-wave $I=0$ $\pi\pi$  scattering length
from twisted mass lattice QCD~\cite{Liu:2016cba}.
M.~Doring {\it et al} extrapolate $\sigma$ resonance
by analyzing the recent results on isoscalar $\pi\pi$ scattering~\cite{Doring:2016bdr}

Although it is quite expensive to implement, the moving wall source technique,
which is originally designed for the center-of-mass frame~\cite{Kuramashi:1993ka,Fukugita:1994ve},
is recognized to compute four-point and three-point functions of
a two-particle scattering with high quality.
Recently we further extended this method to two-particle system
with the non-zero momenta to tentatively investigate
the scalar mesons $\kappa$ and $\sigma$~\cite{Fu:2011xw,Fu:2012gf},
and vector meson $K^\star(892)$ and $\rho$ meson decays~\cite{Fu:2012tj,Fu:2016itp},
along with a few studies of meson-meson scattering~\cite{Fu:2011bz}.
In these works, we confirmed that the moving wall source can calculate
the relevant four-point and three-point correlators with high quality.

It is comparatively cheap to perform lattice studies using staggered fermion
in contrast with those using other lattice discretizations,
consequently, it enables us to carry out lattice examinations
with larger lattice spatial dimension $L$
or smaller quark mass with the fixed and limited computer resources.
For this reason, we study $\sigma$ resonance
on a $40^3\times96$ MILC fine gauge configuration with $N_f=2+1$ flavor
of the ``Asqtad'' improved staggered dynamical sea quarks in this work~\cite{Bernard:2010fr,Bazavov:2009bb}.
We found that the noise-to-signal ratio of three-point function are pretty good,
as compared with four-point function due to the so-called vacuum diagram.

Inspired by the lattice calculations of the hadronic coupling constants of vector mesons
with only three-point functions~\cite{Loft:1988sy,Gottlieb:1983rh,Gottlieb:1985rc,Altmeyer:1995qx},
we will investigate the hadronic coupling constant of a scalar meson strong decay
process $\sigma \to \pi^+ + \pi^-$ (i.e., $g_{\sigma\pi^+\pi^-}$, for short, we called $g_{\sigma\pi\pi}$)
with only the corresponding three-point function,
since, as it is demonstrated later, it can be calculated with high quality in reasonable amounts of computer time.

This needs us to seek a function to parameterize three-point function
in which the hadronic coupling constant $g_{\sigma\pi\pi}$ appears as a parameter.
For this aim, the generating functional method is used to connect the three-point function to the hadronic coupling constant.
Moreover, we should determine the relevant Clebsch-Gordan (CG) coefficients
since the coupling constants are normally denoted as the flavor independent quantities.

\section{The Phenomenological Model}
\label{Sec:Methods}
In this section, the original definitions and notations in Refs.~\cite{Loft:1988sy,Gottlieb:1983rh,Gottlieb:1985rc,Altmeyer:1995qx}
are employed to derive the relationship between the hadronic coupling constant of the scalar meson strong decay
process $S \to A + \pi$ and the three-point function of the scalar-meson field $S(x)$,
together with a pseudoscalar meson field $A(x)$, and a pion meson field $\pi(x)$
\begin{equation}
\label{eq_Gamma3_def}
\Gamma_3 \equiv \langle \tilde{S}({\mathbf q}_S,t_S) \, A(0,t_A) \, \tilde{\pi}({\mathbf q}_\pi, t_\pi) \rangle ,
\end{equation}
where the tilde implies that the relevant field operator is defined
in momentum space, for instance,
$$
\tilde{\pi}({\mathbf q},t) =
\int \frac{d^3\mathbf{x}}{(2\pi)^3} e^{-i{\mathbf q \cdot \mathbf x}} \pi({\mathbf x},t) .
$$
On the lattice, the integral hands over to a sum.

This three-point function can be associated to the three-point function in momentum-space:
\begin{equation}
\label{eq_Gamma3_m}
\Gamma_3 = \int \frac{d^3\mathbf{q}_A}{(2\pi)^3}
          \langle \tilde{S}({\mathbf q}_S,t_S) \, \tilde{A}(\mathbf{q}_A,t_A) \, \tilde{\pi}({\mathbf q}_\pi, t_\pi) \rangle .
\end{equation}
These interacting fields can be renormalized to the asymptotic free fields (i.e., at spatial infinity)
by a field strength renormalization constant $\sqrt{Z}$
so that the single-particle contribution to its propagator has the same behavior
near its pole as the propagator of a free field.
As a result, Eq.~(\ref{eq_Gamma3_m}) can be expressed as
\begin{eqnarray}
\label{eq_Gamma3_inf}
\Gamma_3 \hspace{-0.25cm} &=& \hspace{-0.25cm}\int \frac{d^3\mathbf{q}_A}{(2\pi)^3}
                                \sqrt{Z_S(\mathbf{q}_S)Z_A(\mathbf{q}_A)Z_\pi(\mathbf{q}_\pi)}  \cr
       \hspace{-0.25cm}&\times& \hspace{-0.25cm}\int d^3\mathbf{x}_S   e^{i\mathbf{q}_S   \cdot \mathbf{x}_S}
                \int d^3\mathbf{x}_A   e^{i\mathbf{q}_A   \cdot \mathbf{x}_A}
                \int d^3\mathbf{x}_\pi e^{i\mathbf{q}_\pi \cdot \mathbf{x}_\pi}  \cr
       \hspace{-0.25cm}&\times&\hspace{-0.25cm}\langle {S}_{\mathrm{as}}({\mathbf x}_S,t_S) \, A_{\mathrm{as}}({\mathbf x}_A,t_A) \,
                     {\pi}_{\mathrm{as}}({\mathbf x}_\pi, t_\pi) \rangle ,
\end{eqnarray}
where the subscripts ``as'' in the fields indicate the asymptotic free fields.
In practice, the renormalization constant $Z_S(\mathbf{q}_S)$ in Eq.~(\ref{eq_Gamma3_inf}) can be calculated from the scalar meson amplitude of the operator $S$, i.e.,
\begin{equation}
Z_S(\mathbf{q}_S) = | \langle S, \mathbf{q}_S | \tilde{S}(\mathbf{q}_S;0) | 0 \rangle |^2
\end{equation}
where $|0\rangle$ is the vacuum state. It is similar for the definitions of $Z_A(\mathbf{q}_A)$ and $Z_\pi(\mathbf{q}_\pi)$.

The last term in Eq.~(\ref{eq_Gamma3_inf}) is a typical Euclidean three-point function
met in Lehmann-Symanzik-Zimmermann (LSZ) reduction theory.
One needs a phenomenological model of the strong decay of a scalar meson to measure it.
The process under study is customarily expressed by
the phenomenological interaction term parameterized by a
coupling constant $g_{SA\pi}$. 
The general effective interaction Lagrangian, which parameterizes
the decay of a scalar meson $S$ into a pseudoscalar meson $A$ and a pion meson $\pi$, can be cast as
\begin{equation}
\label{eq_L_def}
{\mathit L}_{\mathrm int}(x) = g_{SA\pi}\mathit{c}_{ijk} S^i(x) A^j(x) \pi^k(x),
\end{equation}
where  $\mathit{c}_{ijk}$ is a Clebsch-Gordan coefficient on the isospin indices $i,j$, and $k$.
Our normalization of  $\mathit{c}_{ijk}$ is chosen so that the Lagrangian manifests like a scalar under
flavor transformations, and for the vertices $\sigma\to\pi\pi$ or $\kappa\to K\pi$,
the kinematic factor $K(\mathbf{q})$ is a constant, i.e.,
\begin{equation}
\label{eq_K_def}
{\mathit K}({\mathbf q}) = 1 .
\end{equation}
It is interesting and important to note that that
kinematic factor $K(\mathbf{q})$ does not depend on the momenta of the participants.

Once the phenomenological model is introduced,
the generating functional method can be applied to solve our problem.
The Euclidean three-point function hinted by this interaction
can be directly estimated from the Feynman path integral through generating functionals, that is
\begin{equation}
\label{eq_gf_inf}
G_3(x_S,x_A,x_\pi) = g\int d^4 x\triangle_S(x-x_S) \triangle_A(x-x_A) \triangle_\pi(x-x_\pi),
\end{equation}
where $\triangle(x-y)$ is the free Feynman propagator between $x$ and $y$,
and  we only consider the tree-level contribution to the generating functionals.

Following the pioneering works in Refs~\cite{Loft:1988sy,Gottlieb:1983rh,Gottlieb:1985rc,Altmeyer:1995qx},
we can readily write the integral in Eq.~(\ref{eq_Gamma3_def}) as the product of three exponentials:
\begin{equation}
\label{eq_Gamma3_1}
\Gamma_3 \equiv f_{SA\pi} \times\int dt e^{-E_S(\mathbf{q}_S)|t_S-t)|} e^{-E_A(\mathbf{q}_A)|t_A-t|} e^{-E_\pi(\mathbf{q}_\pi)|t_\pi-t|} ,
\end{equation}
where we define
\begin{equation}
f_{SA\pi} \equiv g_{SA\pi} \frac{\sqrt{Z_S(\mathbf{q}_S) Z_A(\mathbf{q}_A) Z_\pi(\mathbf{q}_\pi)}}
                                {8E_S(\mathbf{q}_S) E_A(\mathbf{q}_A) E_\pi(\mathbf{q}_\pi)} .
\end{equation}

Assuming large time distances between the operators we can restrict ourselves
to the low lying particle states.
For a fixed source of $A$ meson at time slice $t_A$ and a fixed pion source located at time $t_\pi$,
the sum can be evaluated piecewise for three cases: $t_S> t_\pi > t_A$,
$t_\pi> t_S > t_A$ and $t_A > t_\pi > t_S$.
The following expression can then be derived for $t_S> t_\pi > t_A$:
\begin{eqnarray}
\label{eq_Gamma3_2}
\hspace{-0.7cm}\Gamma_3 \hspace{-0.25cm}&=&\hspace{-0.25cm} f_{SA\pi} \big[ P(E_S, E_\pi+E_A)e^{-E_\pi(t_S-t_\pi)}e^{-E_A(t_S-t_A)}   \cr
\hspace{-0.25cm}&&\hspace{0.3cm}+ P(E_A,   E_S+E_\pi)e^{-E_S(t_S-t_A)}e^{-E_\pi(t_\pi-t_A)} \cr
\hspace{-0.25cm}&&\hspace{0.3cm}+ P(E_\pi, E_S-E_A)e^{-E_S(t_S-t_\pi)}e^{-E_A(t_\pi-t_A)}
\big] ,
\end{eqnarray}
where, to make ease notation, the free lattice particle propagator is denoted as
\begin{equation}
P(\omega,E) \equiv \frac{\sinh(\omega)}{\cosh(\omega)-\cosh(E)}  .
\end{equation}
The other cases can be written down similarly.
Since, in this work, we do not measure the corresponding lattice data,
the explicit form is of no relevance here.

We should remark at this point that, like other staggered hadron operators,
$\sigma$ operator also has the undesired property of
coupling to a state with opposite parity, namely a taste-axial-vector $\eta_A$ meson~\cite{Bernard:2007qf}.
This parity-partner state contributes to the three-point correlators so that
the additional terms should be taken into account in Eq.~(\ref{eq_Gamma3_2}),
as it is done in Ref.~\cite{Altmeyer:1995qx},
which results in the more sophisticated parameterizations.

In principle, Eq.~(\ref{eq_Gamma3_2}) should also include the terms
accounting for the parity partners  of the pion mesons.
However, the pion operators was chosen to be the lattice Goldstone pion,
and the taste of its oscillating parity partners is $\gamma_0$.
This particle (i.e., $\pi_V$ ) is an exotic state since its  $J^{PC} = 0^{+-}$.
Therefore, the parity partner of the pion created by this operator is highly
suppressed such that it can be neglected in the analysis.
(i.e. $Z \sim 0$, as can be noticed in the analysis of pion propagator),
consequently, it is not necessary to be considered in Eq.~(\ref{eq_Gamma3_2}) for the current study.

To estimate the scalar-meson partial width for the aforementioned three-point interaction,
we conveniently begin with the generic two-body decay
rate formula in the center-of-mass frame of the decaying particle
\begin{equation}
d\Gamma_{ijk} = \frac{|{\mathit M}_{ijk}|^2 |\mathbf p_f|}{32\pi^2m_S^2} d\Omega ,
\end{equation}
where the ${\mathit M}_{ijk}$ is a given the matrix element,
$m_S$ is mass of the scalar meson $S$,
and $|\mathbf p_f|$ is the magnitude of either outgoing momentum.

Performing the angular integration,  averaging over the incoming spin states, and
summing over the outgoing spin states, we arrive at
\begin{equation}
\Gamma_{ijk} = \frac{|{\mathit M}_{ijk}|^2 |\mathbf p_f| (2J_A+1)}{8\pi m_S^2 (2J_S+1)}  .
\end{equation}

From the interaction vertex denoted in  Eq.~(\ref{eq_L_def}), we can easily obtain that
the matrix element squared is just
\begin{equation}
|{\mathit M}_{ijk}|^2 =  g_{SA\pi}^2\mathit{c}_{ijk}^2 .
\end{equation}

The final formula for the total decay width can be  expressed as~\cite{Nebreda:2010wv}
\begin{equation}
\Gamma = \frac{g_{SA\pi}^2 |\mathbf p_f| }{8\pi m_S^2} \sum_{ijk} \mathit{c}_{ijk}^2 .
\end{equation}
We should bear in mind that $g_{SA\pi}$ is dimensional.
Like the other dimensional parameters calculated in QCD, it is anticipated
to display dependence on the lattice spacing. 
Fortunately, according to the studies in Ref.~\cite{Nebreda:2010wv},
the scalar coupling constant $g_{SA\pi}$ (i.e., $g_{\sigma\pi\pi}$ or $g_{\kappa K\pi}$)
is a rough constant since it varies pretty slowly as the quark mass changes.
We note that, there is an extra $m_\sigma/2$ in Eq.~(\ref{eq_L_def}) for  $\sigma\to\pi\pi$ in Ref.~\cite{Kleinert:1972ye},
which leads in the coupling  $g_{\sigma\pi\pi}$ to be dimensionless~\cite{Kleinert:1972ye}.

\section{Lattice correlator}
\label{Sec:Correlator}
We have described a detailed procedure to measure $\sigma$ correlator
$\langle 0 | \sigma^\dag(t)  \sigma   (0) | 0 \rangle$~\cite{Fu:2011zzh,Bernard:2007qf,Fu:2012gf}.
To simulate the correct number of quark species,
we use an interpolator with the isospin $I=0$ and $J^{P}=0^{+}$ at source and sink,
$$
{\cal O}(x)  \equiv
\sum_{a, t}
\frac{ \bar u^c_t( x ) u^c_t( x ) +
       \bar d^c_t( x ) d^c_t( x ) }{ \sqrt{2n_t} } ,
$$
where $t$   is the index of taste replica,
      $n_t$ is the number of taste replicas, and
      $c$   is color index.
After carrying out the Wick contractions of fermion fields,
and summing over the taste and color indexes,
we obtain the time slice correlator $C(t)$ with momentum ${\mathbf p}$
\begin{eqnarray}
\label{EQ:CBB}
C(\mathbf{p},t) \hspace{-0.2cm}&=&\hspace{-0.2cm}
-\frac{1}{2} \sum_{ \mathbf{x} }
e^{i \mathbf{p} \cdot \mathbf{x} }
\langle
\mbox{Tr} M^{-1} (\mathbf{x},t; \mathbf{x},t)
\mbox{Tr} M^{-1} ({\bf 0},0;{\bf 0},0)
\rangle \cr
\hspace{-0.2cm}&+&\hspace{-0.2cm}
\sum_{ \mathbf{x} }(-)^x e^{i \mathbf{p}\cdot \mathbf{x} }
\langle \mbox{Tr}
[ M^{-1}     (\mathbf{x},t;{\bf 0},0)
  M^{-1^\dag} \mathbf{x},t;{\bf 0},0) ]
\rangle, \nonumber
\end{eqnarray}
where $M$ is the light quark Dirac operator, and the first and second  terms are the
quark-line disconnected and connected contributions, respectively~\cite{Fu:2011zzh,Bernard:2007qf,Fu:2012gf}.

Like other staggered hadron operators,
the $\sigma$ operator also has the undesired oscillating term with opposite parity,
namely a taste-axial-vector $\eta_A$ meson~\cite{Fu:2011zzh,Bernard:2007qf,Fu:2012gf}.
In practice, we take one mass with each parity~\cite{Fu:2011zzh,Bernard:2007qf,Fu:2012gf}.
Then, the $\sigma$ correlator was fit to
\begin{equation}
\label{eq:kfit}
  C_{\sigma}(t)  = Z_{\sigma}e^{-m_{\sigma}t} +
  Z_{\eta_A}(-)^t e^{-M_{\eta_A}t} + (t \rightarrow N_t-t),
\end{equation}
where $Z_{\sigma}$ and $Z_{\eta_A}$ are two overlap factors.

We should bear in mind that, for the staggered Kogut-Susskind quark action,
the $\sigma$ interpolator couples to various tastes
as we examined in our previous studies for scalar $a_0$, $\sigma$
and $\kappa$ mesons~\cite{Fu:2011zzh,Bernard:2007qf,Fu:2012gf},
where we investigated the bubble contribution, and found that
it dominates the correlator at large time distance.
Thus, we should remove this term from $\sigma$ correlator.

It is well-known that the four-point and three-point functions
are very difficult to calculate,  and the so-called stochastic source method,
or its variants: the distillation method, etc. are normally used to compute~\cite{Peardon:2009gh}.
Although it is quite expensive to implement, the moving wall source technique
is used to compute three-point functions in this work.

To avoid the Fierz rearrangement of the quark lines, we choose $t_1 =0, t_2=1$, and $t_3=t$
for the $\sigma \to \pi\pi$ three-point function~\cite{Fukugita:1994ve}.
The quark line diagrams contributing to $\sigma \to \pi\pi$
three-point correlation function is displayed in Fig.~\ref{fig:3diagram}.
The calculation of the $\sigma \to \pi\pi$ three-point function is quite difficult.
In practice, we employ an up-antiquark source with $1$ on each lattice site ${\mathbf{x}}$
for pion creation operator, and an up-quark source with $e^{i{\mathbf{p}}\cdot{\mathbf{x}}}$
on each lattice site ${\mathbf{x}}$ for pion creation operator~\cite{Fu:2012gf}.

\begin{center}
\includegraphics[width=4.25cm,clip]{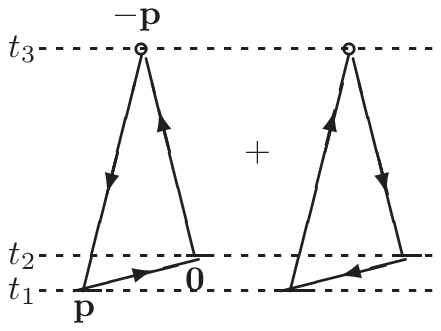}
\figcaption{
\label{fig:3diagram}
Quark line contraction diagrams of $\sigma \to \pi\pi$.
Short bars represent the wall sources. Open circles stand for the sinks of local pion operator.
}
\end{center}

It should be worthwhile to stress that the imaginary part of the second diagram
for  $\sigma \to \pi\pi$ has same magnitude but with minus sign,
as that of the first diagram.
As a consequence, three-point diagrams are purely real,
and only one quark-line diagram is required to calculate. 
It is interesting and important to note that the value in Eq.~(\ref{eq_Gamma3_2})
is also purely real, as expected.

We then write the first diagrams for the $\sigma \to \pi\pi$
quark-line diagrams in Fig.~\ref{fig:3diagram}
in terms of the light quark propagators $G$,
\begin{eqnarray}
\label{eq:dcr3}
\hspace{-0.5cm}C_{\sigma \to \pi\pi} ({\mathbf p},t_3,t_2,t_1) &=& \cr
&&\hspace{-3.5cm}\mbox{Re} \sum_{ {\mathbf{x}}_3, {\mathbf{x}}_1}
  e^{ i {\mathbf p} \cdot {\mathbf{x}}_3 }
\left\langle \mbox{Tr}
[G_{t_1}({\mathbf{x}}_3, t_3) G_{t_2}^{\dag}({\mathbf{x}}_3, t_3)
 G_{t_2}^{\dag}({\mathbf{x}}_1, t_1) ] \right\rangle,
\end{eqnarray}
where the trace is taken over color index and  Dirac matrix
is used as an interpolator for $i$th meson:
the $\gamma_5$ for the pseudoscalar meson and $1$ for $\sigma$ meson.

\section{Lattice calculation}
\label{sec:latticeCal}
In this work, we used $400$ MILC $40^3 \times 96$ gauge configurations with $2+1$ flavors
of the Asqtad-improved staggered fermions,
where bare quark masses $am_{ud}/am_s = 0.0031/0.031$
and bare gauge coupling $10/g^2 = 7.08$~\cite{Bernard:2010fr,Bazavov:2009bb,Bernard:2001av,Aubin:2004wf}.
The lattice spacing $a$ is about $0.09$~fm.
The masses of $u$ and $d$ quarks are degenerate.
All the gauge configurations were gauge fixed to the Coulomb gauge
before calculating the quark propagators.
The standard conjugate gradient method is utilized to acquire
the required matrix elements of inverse Dirac fermion matrices,
and the conjugate gradient residual is selected to be $1.0\times10^{-5}$,
which is generally smaller than that of generating the MILC gauge
configurations~\cite{Bernard:2001av}.
Moreover, all the numerical calculations are evaluated in double precision
to avoid the potential roundoff errors.

In the calculation of the $\sigma$ correlator,
$$
\sigma(t)=\frac{1}{T}\sum_{t_s}\langle\sigma^\dag(t+t_s) \sigma(t_s)\rangle ,
$$
we average all the possible correlators.
One thing we must stress is that we use $Z_2$ noisy estimators
based on the random color fields to measure the disconnected contribution
of sigma correlator~\cite{Fu:2011zzh}.
Using the standard discussed in ref.~\cite{Muroya:2001yp},
we determine that $1000$ noise $Z_2$ sources are sufficiently reliable
to measure the disconnected part.
Fitting $\sigma$ correlator with Eq.~(\ref{eq:kfit}), we can obtain the sigma mass $m_\sigma$
and two overlap amplitudes: $Z_{\eta_A}$ and $Z_{\sigma}$,
which will be subsequently plugged into Eq.~(\ref{eq_Gamma3_2}) to estimate the three-point functions.

We compute the $\sigma\to\pi\pi$ three-point functions on all the time slices,
and explicitly combine the results from all the time slices $T$;
namely, the $\sigma\to\pi\pi$ three-point correlator $C_{\sigma\to\pi\pi}(t)$ is measured through
\begin{eqnarray}
 C_{\sigma\to\pi\pi}(t) &=& \frac{1}{T}\sum_{t_s}^{T}
\left\langle \sigma(t+t_s)(\pi\pi)^\dag(t_s)\right\rangle ,
\nonumber
\end{eqnarray}
After averaging the propagators over all the $T$ values,
the statistics are found to be remarkably improved.

According to the discussion in Appendix of Ref.~\cite{Fu:2016itp},
the noise-to-signal ratio of $\sigma$ correlator and $\sigma\to\pi\pi$ correlator
are improved as approximately  $\propto \frac{1}{\sqrt{N_{\rm slice} L^3}}$,
where $L$ is lattice spatial dimension and $N_{\rm slice}$ is the
number of the time slices calculated the propagators for each of the gauge configurations.
In this work, we use the lattice ensembles with relatively large $L$
and sum $\sigma$ correlator over all the time slices
and the $\sigma\to\pi\pi$ correlator over all the time slices;
consequently, it is natural that the signals of the correlators
should be significantly improved.
Admittedly, the most efficient way to improve the relevant noise-to-signal ratio
is to use finer gauge configurations or anisotropic gauge configurations~\cite{Wilson:2015dqa}.

We measure two-point pion correlators with the zero and nonezero
momenta ($\mathbf{0}$ and $\mathbf{p}$) as well,
\begin{eqnarray}
\label{eq:pi_cor_PW_k000}
C_\pi({\mathbf 0}, t) &=& \frac{1}{T}\sum_{t_s=0}^{T-1}
\langle 0|\pi^\dag ({\mathbf 0}, t+t_s) W_\pi({\mathbf 0}, t_s) |0\rangle, \\
\label{eq:pi_cor_PW_k100}
C_\pi({\mathbf p}, t) &=& \frac{1}{T}\sum_{t_s=0}^{T-1}
\langle 0|\pi^\dag ({\mathbf p}, t+t_s) W_\pi({\mathbf p}, t_s) |0\rangle,
\end{eqnarray}
where $\pi$ is the pion point-source operator
and $W_\pi$ is the pion wall-source operator~\cite{Bernard:2001av,Aubin:2004wf}.
It is worth noting that the summations over all the time slices for $\pi$ propagators
guarantee the extraction of $\pi$ mass with high precision.

Disregarding the contributions from the excited states,
the pion mass $m_\pi$ and energy $E_\pi({\mathbf p})$
can be robustly extracted at large $t$ from two-point pion correlators~(\ref{eq:pi_cor_PW_k000})
and (\ref{eq:pi_cor_PW_k100}), respectively~\cite{Bazavov:2009bb},
\begin{eqnarray}
\label{eq:pi_fit_PW_k000}
\hspace{-0.6cm} C_\pi({\mathbf 0}, t) &=& Z_\pi(\mathbf{0}) \left[e^{-m_\pi t}+e^{-m_\pi(T-t)}\right] +\cdots, \\
\label{eq:pi_fit_PW_k100}
\hspace{-0.6cm} C_\pi({\mathbf p}, t) &=& Z_\pi(\mathbf{p})
\left[e^{-E_\pi({\mathbf p}) t}+e^{-E_\pi({\mathbf p})(T-t)}\right] + \cdots,
\end{eqnarray}
where the ellipses show the oscillating parity partners,
and $Z_\pi(\mathbf{0})$ and $Z_\pi(\mathbf{p})$ are two overlapping amplitudes,
which will be subsequently plugged into Eq.~(\ref{eq_Gamma3_2})
to estimate three-point correlation functions.

\section{Simulations and results}
\label{Sec:Results}

The valence $u/d$ quark masses are set to its dynamical quark masses, 
while valence strange quark is fixed to its physical mass,
which was determined by MILC Collaboration~\cite{Bernard:2010fr}.
In the usual manner, we extracted $\pi$, $K$,
and fictitious $s\bar{s}$ masses~\cite{Bernard:2001av}, which are used to evaluate
bubble contribution to $\sigma$ correlators~\cite{Fu:2011zzh,Bernard:2007qf,Fu:2012gf},
where three low energy couplings ($\mu$, $\delta_A$, and $\delta_V$)
are fixed to MILC-determinated values~\cite{Aubin:2004fs}.
After neatly removing the unwanted bubble terms from $\sigma$ propagators,
the remaining $\sigma$ propagators have a clean information,
we then fit them with the physical model in Eq.~(\ref{eq:kfit}).
The extracted meson masses gave $m_\pi/m_\sigma \approx 0.38 < 0.5$,  
ensuring that the physical kinematics for the $\sigma$-meson decay is satisfied.
In this work, we directly quote the lattice parameters,
which are professionally determined by MILC Collaboration~\cite{Bernard:2010fr,Bernard:2001av,Aubin:2004fs}.

First of all, we adopt the natural choice of having $\sigma$ meson at rest and two pions at rest.
Moreover, we also measure the case with two pions at the nonzero momentum
since we found that the signal of the Goldstone pion propagator with the nonzero momentum
are also stable with our particular choice of kinematics.
The time slice $t_1$ of the pion has been fixed to $t_1 = 1$,
another pion was located at $t_\pi = 0$,
and we evaluated the correlation function for all times $t$ of the $\sigma$ meson.
In this work, $\pi$ meson was given the minimal lattice momentum ${\mathbf p}_\pi = (2\pi/L){\mathbf e}_z$.

Figure~\ref{fig:3pointC} shows our computed three-point function on the lattice
as a function of the temporal location of the scalar meson
where the scalar $\sigma$ meson is at rest.
Octagon red one indicates the three-point function with two pions at rest, and
square blue one shows three-point function with two pions
at momenta ${\mathbf p}_\pi = (2\pi/L){\mathbf e}_z$ and
$-(2\pi/L){\mathbf e}_z$,respectively.
As expected, we observe the very clear signal.
This situation is quite different from that reported for vector $\rho$ meson,
where far from the source $\Gamma_3$ is consistent with zero~\cite{Loft:1988sy}.

It is interesting and important to note that the oscillating parity partners
in three-point functions are not clear in Fig.~\ref{fig:3pointC}.
This is easy to understand, since the parity partner of lattice Goldstone pion
corresponds to an exotic state and is highly suppressed so that it can be neglected in the analysis.
As a consequence, the MILC Collaboration usually
adopts a fit of type  ``1,0'' in the fit of pion mass£¬
which means that the oscillating parity partner is not included~\cite{Aubin:2004wf}.
We should remark at this point that, the $\sigma$ operator also has the undesired property of
coupling to a state with opposite parity, namely a the taste-axial-vector $\eta_A$ meson~\cite{Bernard:2007qf}.
However, the sigma mass $m_\sigma$ is much smaller than that of $\eta_A$ meson
for the MILC lattice ensemble used in the present work,
so it is highly suppressed so that it  can be neglected in the analysis.

\begin{center}
\includegraphics[width=8.cm,clip]{gspp_31.eps}
\figcaption{
\label{fig:3pointC}
Real parts of $\sigma\to\pi\pi$ three-point function with $\sigma$ meson at rest.
Occasional points with negative central values are not displayed.
}
\end{center}

In principle, we can use the above information to get the fitting value of
the coupling constant $g_{\sigma\pi\pi}$.
To get more information, the three-point functions were also generated
by giving the $\sigma$-meson a momentum $\mathbf{p}_\sigma$
and varying $\sigma$-meson time-slice location $t$.
In this work, we chose to put the same spatial momentum ${\mathbf p}$ on the $\sigma$-meson and one pion meson,
and another pion was set to ${\mathbf 0}$.
Our signal is much more stable with our particular choice of kinematics.
The time slice $t_1$ of the pion has been fixed to $t_1 = 1$,
another pion was located at $t_\pi = 0$,
and we evaluated the correlation function for all times $t$ of the $\sigma$ meson.
The $\sigma$ meson was given the minimal lattice momentum ${\mathbf p}_\sigma = (2\pi/L){\mathbf e}_z$ in the $z$ direction,
and we also measure at ${\mathbf p}_\sigma =(2\pi/L)({\mathbf e}_y+{\mathbf e}_z)$,
$(2\pi/L)({\mathbf e}_x+{\mathbf e}_y+{\mathbf e}_z$), and $(2\pi/L)(2{\mathbf e}_z$)
(i.e., ${\mathbf p}_\sigma$ = $[0,0,1]$, $[0,1,1]$, $[1,1,1]$, $[0,0,2]$).

Figure~\ref{fig:3pointC2} shows our lattice-measured three-point functions
with $\sigma$ meson at above-mentioned four momenta
as a function of the temporal location of the sigma meson.
As expected, we observe a very clear signal, and it is interesting to note that
the oscillating behavior generally contributes in relatively larger quantities for the higher momenta.
This is easy to comprehend since the sigma energy $E_\sigma$ is more and more closer
with $\eta_A$ meson energy $E_{\eta_A}$ for the higher momenta.

We are now in a position to discuss the flavor quantum numbers of the states we are investigating,
since in lattice calculation, the coupling of any $\bar{u}u$ pair to a meson is assumed to be just unity.
The $\sigma$, $\pi^+$ and  $\pi^-$ wave functions are $\frac{1}{\sqrt{2}}(u\bar{u}+d\bar{d})$, $\bar{d}u$, $\bar{u}d$, respectively.
As a consequence, there exists a factor of $2$
since the $\bar{q}q$ pair which pops out of the vacuum
can be either a $\bar{u}u$ pair or a $\bar{d}d$ pair when the $\sigma$ decays~\cite{Loft:1988sy}; therefore
\begin{equation}
g_{\sigma\pi\pi} = \sqrt{2} \, g_{\mathrm{lattice}} \,.
\end{equation}

\begin{center}
\includegraphics[width=8.cm,clip]{gspp_0031.eps}
\figcaption{
\label{fig:3pointC2}
Real parts of $\sigma\to\pi\pi$ three point function with $\sigma$ meson at momentum ${\mathbf p}$.
Occasional points with negative central values are not displayed.
}
\end{center}

Now we are ready to determine the $g_{\sigma\pi\pi}$ coupling
constant from the numerical data which are shown in Fig.
~\ref{fig:3pointC} and Fig.~\ref{fig:3pointC2}.
All these constants together with the masses and energies of the different particle states
needed as inputs to Eq.(\ref{eq_Gamma3_2}) were extracted from the analysis of two-point functions,
and then used to obtain the theoretical form of three-point function.
The coupling constant $g_{\sigma\pi\pi}$ is determined by fitting this function to
the lattice-measured three-point functions, discarding various choices of time slices.

The three-point functions were measured with $\sigma$ meson at five momenta
$\mathbf{p}$=$(0,0,0)$, $(0,0,1)$, $(0,1,1)$, $(1,1,1)$, and $(0,0,2)$.
All six correlators (two in $\mathbf{p}$=$(0,0,0)$) were then simultaneously fitted to
the physical model in Eq.(\ref{eq_Gamma3_2}) for only one fitting parameter $g_{\sigma\pi\pi}$.
We find for the local  $\sigma$ operator
\begin{equation}
g_{\sigma\pi\pi} = 2.71\pm0.42 \,\, {\mathrm{GeV}} .
\end{equation}
This is in reasonable agreement with the recent analytic predictions from residue at complex pole,
which are listed in Table~\ref{tab:g_spp}, together with our former lattice result~\cite{Fu:2012gf}.
This is also in fair accordance with Hadron Spectrum Collaboration's lattice result~\cite{Briceno:2016mjc}.
The agreement is fairly reasonable, taking into account that,
according to the studies in Ref.~\cite{Nebreda:2010wv},
the scalar coupling constant $g_{SA\pi}$ (i.e., $g_{\sigma\pi\pi}$ or $g_{\kappa K\pi}$)
is a rough constant since it varies pretty slowly as the quark mass changes.
Moreover, the Hadron Spectrum Collaboration found that
the coupling constant $g_{\sigma\pi\pi}$ is approximately independent of quark mass~\cite{Briceno:2016mjc}.

We should remark at this point that $g_{SA\pi}$ is not dimensionless,
it is expected to show dependence on the lattice spacing.
Since we only work on a MILC lattice ensemble, our obtained  $g_{SA\pi}$ is not physical.
More sophisticated evaluation should be carried out at several lattice ensembles,
and discuss the mass dependence.

\begin{center}
\tabcaption{
\label{tab:g_spp}
Other recent determinations of the $\sigma$ meson coupling to two pions ($g_{\sigma\pi\pi}$)
using some form of analytic properties or data constrained by Roy equations and chiral symmetry.
}
\footnotesize
\begin{tabular*}{85mm}{@{\extracolsep{\fill}}lc}
\toprule
$\mathrm{Reference}$ & $|g_{\sigma\pi\pi}|$ (GeV)     \\
\hline
Pelaez~\cite{Pelaez:2015qba}                                    & $3.45^{+0.25}_{-0.22} $    \\
Masjuan, Ruiz de Elvira and Sanz-Cillero~\cite{Masjuan:2014psa} & $3.8\pm0.4$  \\
Oller~\cite{Oller:2003vf}                                       & $2.97\pm0.05$   \\
Kaminski, Mennessier and Narison~\cite{Kaminski:2009qg}         & $2.2$   \\
Mennessier, Narison and Wang~\cite{Mennessier:2010xg}           & $2.65\pm0.1$   \\
Garcia-Martin et al.~\cite{GarciaMartin:2011jx}                 & $3.59^{+0.11}_{-0.13}$   \\
Pelaez and Rios~\cite{Pelaez:2010fj} (fit D)                    & $3.5$   \\
Narison~\cite{Narison:2005wc}                                   & $5.3\pm1.8$   \\
Nebreda and Pelaez~\cite{Nebreda:2010wv}                        & $2.86$  \\
\hline
Fu~\cite{Fu:2012gf}                                             & $2.69\pm0.44$\\
This work                                                       & $2.71\pm0.42$\\
\bottomrule
\end{tabular*}
\end{center}

\section{Summary and outlook}
\label{Sec:Conclusions}
In this work, we discuss that the hadronic coupling constants for the scalar-meson
strong decays $S\to A\pi$ are extracted from the lattice three-point function.
And we report an exploratory lattice investigation of the hadronic coupling constants $g_{\sigma\pi\pi}$ for the hadronic decays $\sigma\to\pi\pi$
only using the appropriate three-point function,
which are evaluated by the moving-wall source technique with a pretty good noise-to-signal ratio.
These simulations are carried out on a $40^3\times96$ MILC fine gauge configuration
with $N_f=2+1$ flavor of the asqtad-improved staggered dynamical sea quarks
at $ m_\pi / m_\sigma \approx 0.38$ and the lattice spacing $a \approx 0.09$ fm.

Our estimated value for this given MILC fine lattice gauge ensemble $g_{\sigma\pi\pi}=2.71(42))$GeV,
which can be reasonably compared to the recent analytic predictions from residue at complex pole,
which are summarized  in Table~\ref{tab:g_spp}, along with our former tentative lattice result~\cite{Fu:2012gf}.
The most important outcome of the lattice calculation of $g_{\sigma\pi\pi}$
exhibits that the lattice study of the scalar meson decay processes
can be carried out in reasonable amounts of time on presently limited available computers.

We should remark at this point that $g_{\sigma\pi\pi}$ is dimensional,
it is expected to show dependence on lattice spacing.
Since we only work on a MILC lattice ensemble, strictly speaking, our obtained  $g_{\sigma\pi\pi}$ is not the physical one,
we can not directly compare with other data.
More sophisticated one should be carried out at several lattice ensembles,
and discuss the mass dependence on lattice spacing.
It will be interesting to see whether this expectation is
borne out in numerical QCD simulations, especially at smaller lattice spacing.

Moreover, according to the empirical discussion in the Appendix of Ref.~\cite{Fu:2016itp},
to improve the relevant noise-to-signal ratio, we should use very fine gauge
configurations or the lattice ensembles with larger lattice spatial dimensions $L$.
For this reason, we are beginning a series of numerical simulations
with the super-fine or ultra-fine MILC lattice ensembles.

Furthermore, admittedly, thee method described in this work cannot obtain
another resonance parameter: the $\sigma$ resonance mass.
To achieve the $\sigma$ resonance mass, we must calculate
the $I=0$ $\pi\pi$ scattering with the careful treatment of vacuum diagrams,
as it done in Ref.~\cite{Fu:2012gf}.
Nonetheless, the reliable evaluation of vacuum diagram needs more lattice or more finer gauge configurations.

All of these open questions are beyond the scope of this paper since this
will demand a huge amount of computing allocations.
We postpone and reserve  these expensive tasks for our future lattice study.
We will enthusiastically appeal for all the possible computational resources
to carry out these challenging tasks.

\acknowledgments{
We deeply appreciate MILC for using MILC gauge configurations and codes.
We sincerely thank Carleton DeTar for inculcating us in the necessary knowledge for this work.
We especially thank Eulogio Oset and Michael Doring for their enlightening comments.
The authors express respect to Han-qing Zheng, Geng Liseng, Liu Chuan, and Chen Ying
for reading this manuscript or providing useful comments.
We cordially express our boundless gratitude to Hou qing, He Yan and Fujun Gou's vigorous support.
We also express gratitude to the Institute of Nuclear Science and Technology, Sichuan
University, and the Chengdu Jiaxiang Foreign Language
School, from which computer resources were furnished.
Numerical calculations for this work were carried out at both the PowerLeader
Clusters and the AMAX, CENTOS, HP, and ThinkServer workstations.
}

\end{multicols}
\vspace{-1mm}
\centerline{\rule{80mm}{0.1pt}}
\vspace{2mm}

\begin{multicols}{2}

\end{multicols}

\clearpage

\end{CJK*}
\end{document}